\documentclass[a4paper]{proceedings}

\usepackage{graphicx}
\usepackage{subfig}
\usepackage{epstopdf}

\begin{document}

\title{E-sail test payload of ESTCube-1 nanosatellite}

\author{
Jouni Envall\textsuperscript{a,b}\corauthref{Corresponding author, jouni.envall@fmi.fi},
Pekka Janhunen\textsuperscript{a},
Petri Toivanen\textsuperscript{a},
Mihkel Pajusalu\textsuperscript{c},
Erik Ilbis\textsuperscript{c},
Jaanus Kalde\textsuperscript{c},
Matis Averin\textsuperscript{c},
Henri Kuuste\textsuperscript{c},
Kaspars Laizans\textsuperscript{b,c},
Viljo Allik\textsuperscript{b},
Timo Rauhala\textsuperscript{d},
Henri Sepp\"anen\textsuperscript{d},
Sergiy Kiprich\textsuperscript{d},
Jukka Ukkonen\textsuperscript{d},
Edward Haeggstr\"om\textsuperscript{d},
Taneli Kalvas\textsuperscript{e},
Olli Tarvainen\textsuperscript{e},
Janne Kauppinen\textsuperscript{e},
Antti Nuottaj\"arvi\textsuperscript{e,f},
Hannu Koivisto\textsuperscript{e}
}

\address{\textsuperscript{a} Finnish Meteorological Institute, P.O. Box 503, FI-00101 Helsinki, Finland}

\address{\textsuperscript{b} Tartu Observatory, 61602, T\~oravere, Tartu County, Estonia}

\address{\textsuperscript{c} University of Tartu, Faculty of Science and Technology, Institute of Physics, T\"{a}he 4-111, 51010, Tartu, Estonia}

\address{\textsuperscript{d} University of Helsinki, Department of Physics, Electronics Research Laboratory, P.O. Box 64, FI-00014, Finland}

\address{\textsuperscript{e} University of Jyv\"askyl\"a, Department of Physics, FI-40014 Jyv\"askyl\"a, Finland}

\address{\textsuperscript{f} Current affiliation: Tampere University of Technology, Department of Chemistry and Bioengineering, FI-33720 Tampere, Finland}


\abstract{The scientific mission of ESTCube-1, launched in May 2013,
  is to measure the Electric solar wind sail (E-sail) force in
  orbit. The experiment is planned to push forward the development of
  E-sail, a propulsion method recently invented at the Finnish
  Meteorological Institute. E-sail is based on extracting momentum
  from the solar wind plasma flow by using long thin electrically
  charged tethers. ESTCube-1 is equipped with one such tether,
  together with hardware capable of deploying and charging it. At the
  orbital altitude of ESTCube-1 (660--680~km) there is no solar wind
  present. Instead, ESTCube-1 shall observe the interaction between
  the charged tether and the ionospheric plasma. The ESTCube-1 payload
  uses a 10-meter, partly two-filament E-sail tether and a motorized
  reel on which it is stored. The tether shall be deployed from a
  spinning satellite with the help of centrifugal force. An additional
  mass is added at the tip of the tether to assist with the
  deployment. During E-sail experiment the tether shall be charged to
  500~V potential. Both positive and negative voltages shall be
  experimented with. The voltage is provided by a dedicated high
  voltage source and delivered to the tether through a slip ring
  contact. When the negative voltage is applied to the tether, the
  satellite body is expected to attract electron flow capable of
  compensating for the ion flow, which runs to the tether from the
  surrounding plasma. With the positive voltage applied, onboard cold
  cathode electron guns are used to remove excess electrons to
  maintain the positive voltage of the tether. In this paper we
  present the design and structure of the tether payload of
  ESTCube-1.}


\maketitle

\section{Introduction}

The scientific mission of ESTCube-1, the first Estonian satellite, is
to perform an on-orbit test of the Electric solar wind sail (E-sail)
concept. The E-sail is a propulsion innovation made at the Finnish
Meteorological Institute (FMI) in 2006. \cite{Janhunen2007AnnGeophys}
Thrust is produced by harnessing the momentum of charged solar wind
particles by using long, thin, and electrically charged conducting
tethers. A fullscale E-sail spacecraft is planned to have e.g. 100
tethers, each 20 km long. The sail is kept open with the help of
centrifugal force, the spacecraft must therefore be kept in rotating
motion. Once operational, E-sail technology is expected to
revolutionize the space travel within our solar
system. \cite{JanhunenProcEstAcadSci2014, Janhunen2008JBIS,
  Mengali2009CelMechDynAstron, Janhunen2009AnnGeophysNegative,
  Toivanen2009, Quarta2010HazardAsteroids, Janhunen2010RevSci,
  Merikallio2010Asteroidit}

The experiment conducted by ESTCube-1 marks the beginning of space
test era for the E-sail. \cite{Silver2014} The ESTCube-1 payload (PL)
includes one E-sail tether, deployable to $\sim$10-meter length. The
tether is stored on a motorized reel and it can be reeled out upon
request from the ground. The principle of tether deployment is similar
to the proposed principle of larger E-sails, i.e. the satellite is
spun around its axis of maximum moment of inertia and the exiting
tether is streched by the effect of centrifugal force. This force is
enhanced by placing a small end mass at the tip of the
tether. Successful deployment of the tether is verified by observing a
noticeable drop in the satellite spin rate. In addition, a visual
verification shall be obtained by imaging the tether and its end mass
with the onboard camera during the tether deployment. Once deployed,
the tether shall be charged with a voltage of 500 V. Both positive and
negative voltages, corresponding to positively and negatively charged
E-sails, respectively, shall be tested. Because of its orbit (a
low-Earth orbit with 660--680 km altitude) ESTCube-1 is not influenced
by solar wind. Instead, the charged tether shall interact with the
ionospheric plasma, through which the satellite is traveling with its
orbital speed of 7.5 km/s. When the on and off cycles of the tether
voltage are correctly synchronized to the satellite's spin rate, the
E-sail effect between the tether and the plasma can be observed as a
cumulative change in the spin rate. The change can be chosen to be in
either direction, depending on the on/off cycling.

In this paper we give an overall technical description of the onboard
apparatus which shall be used to perform the E-sail experiments. The
satellite was launched into orbit on May 7th 2013 (UTC), but at the
time of writing it had not yet entered the tether experiment phase of
its mission.

\section{Apparatus}

\begin{figure}
\centerline{\includegraphics[width=0.6\columnwidth]{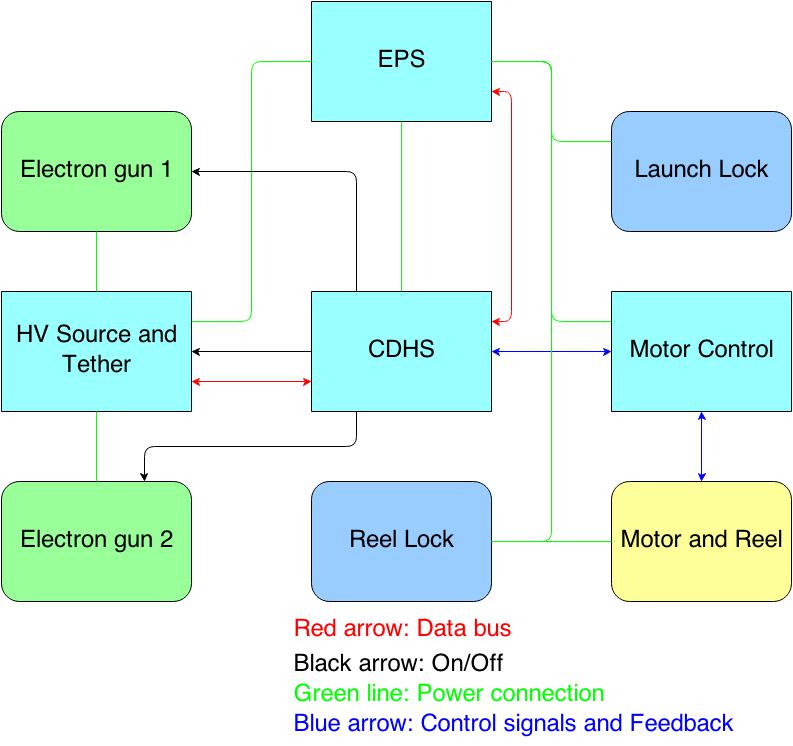}}
\caption{
Block diagram depicting the data and power interactions between PL parts, CDHS and EPS.
}
\label{FigDiagram}
\end{figure}

\begin{figure}
\centerline{\includegraphics[width=0.6\columnwidth]{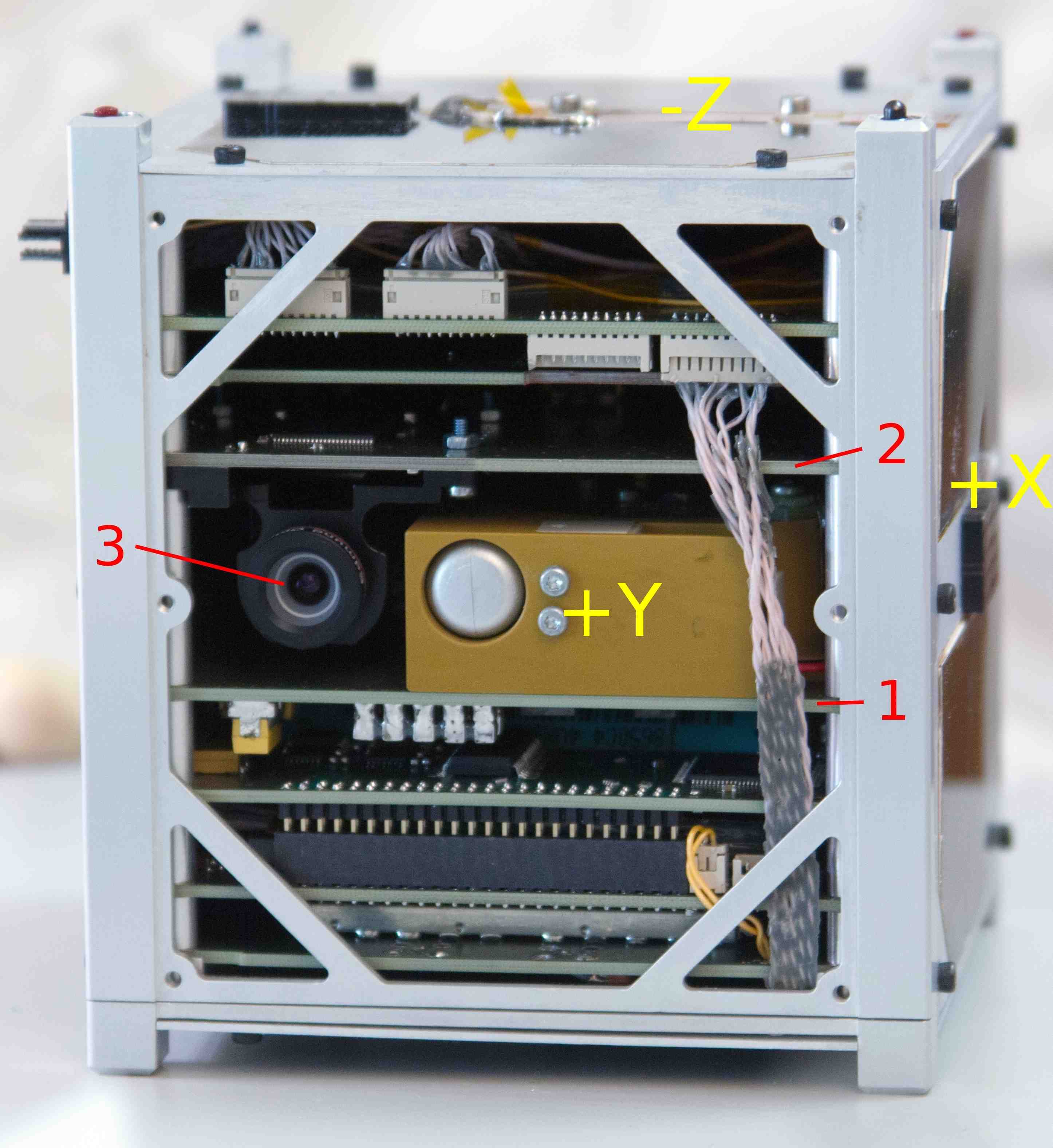}}
\caption{
Assembled satellite with side panels removed. Labels: (1) Motor board,
(2) HV board, (3) onboard camera. The yellow labels denote the
directions of X, Y and Z axes of the satellite's coordinate system.
}
\label{FigSatOpen}
\end{figure}

The PL apparatus is divided on two separate circuit boards, the motor
board and the high voltage (HV) board. Figure~\ref{FigDiagram} depicts
the internal connections of PL, as well as the interactions between PL
and the Command and Data Handling System (CDHS) and the Electrical
Power System (EPS) of the satellite. Figure~\ref{FigSatOpen} shows how
the circuit boards are fitted into the assembled satellite. The motor
board is responsible for the tether deployment. It is where the tether
is stored prior to the deployment and it also includes the hardware
and electronics needed to reel out the tether. HV board provides the
$\pm$500~V voltages needed during the E-sail experiment. In addition
to the circuit boards, two cold cathode electron guns are attached on
$\pm$Z sides of the satellite, one on each side.

The HV board does not have its own microcontroller unit (MCU). The HV
operations are directly operated and commanded by CDHS. CDHS is able
to switch the high voltage on or off and select which of the electron
guns, if any, are operational. These operations are commanded via
simple digital on/off signals. In return the HV board provides CDHS
the signals for current measurement, see section \ref{SecHV} for
details. Both tether current and the (electron gun) anode current are
measured throughout the E-sail experiment. The former provides
scientific data needed for the full understanding of the E-sail
phenomenon while the latter provides information about the performance
of the electron guns.

The motor board includes a MCU, but the communication between the
motor board and the CDHS is nevertheless handled via low level on/off
signals. CDHS commands the motor on or off and in return it gets a
pulse sequence, corresponding to the amount which the motor has
turned.

\subsection{Tether}
\begin{figure}
  \centering
  \subfloat[][Photograph of tether. Labels: (1) Basewire, (2) Loop wire. Distance between bonds is 10 mm.]{\label{FigTetherPhoto}\includegraphics[width=0.5\columnwidth]{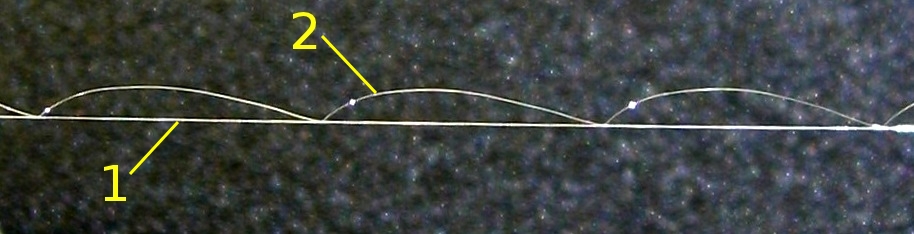}}\\
  \subfloat[][Scanning Electron Microscope (SEM) image of an ultrasonic wire-to-wire bond.]{\label{FigBond}\includegraphics[width=0.5\columnwidth]{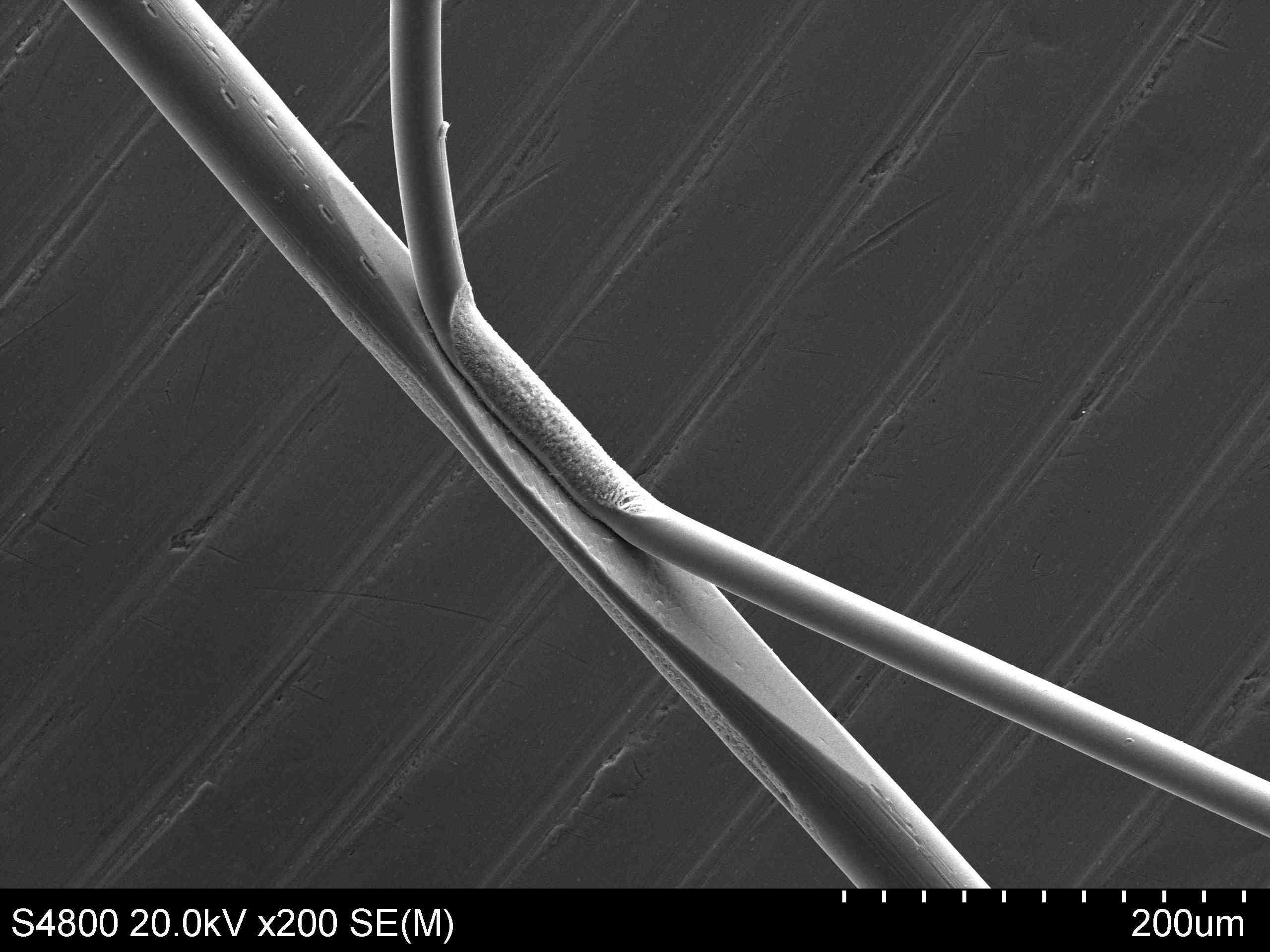}}
  \caption{Two-filament Heytether.}
  \label{FigTether}
\end{figure}

The tether is one of the core technologies of E-sail development. The
tether manufacture is being carried out by the Electronics Research
Laboratory at the University of Helsinki (UH). Currently E-sail
tethers are made of 25--50~$\mu$m aluminum wires, which are
ultrasonically bonded together to form multifilament tethers. UH has
developed a specific ultrasonic wire-to-wire bonding technology to
achieve this
goal.\cite{Seppaenen2011WireToWire,SeppanenOneKilometer2013} The
reason for using the multifilament design is to make the tethers more
resistant against micrometeoroid bombardment. The tether onboard
ESTCube-1 consists of two parts; a 3.5-meter two-filament Heytether
part and a 16-meter single filament part. The physical properties of
the Heytether part enable proper demonstration of E-sail tether
outreeling, and it also provides protection against micrometeoroids
for the outer end of the tether. The single filament part enables the
overall length of the tether to be lengthened up to 19.5
meters. Nominally the ESTCube-1 mission is designed to operate a
10-meter tether. Figure~\ref{FigTether} illustrates the Heytether
structure. It consists of a 50-$\mu$m basewire and a 25-$\mu$m loop
wire.

\subsection{Tether reel and isolation}
\begin{figure}
\centerline{\includegraphics[width=0.6\columnwidth]{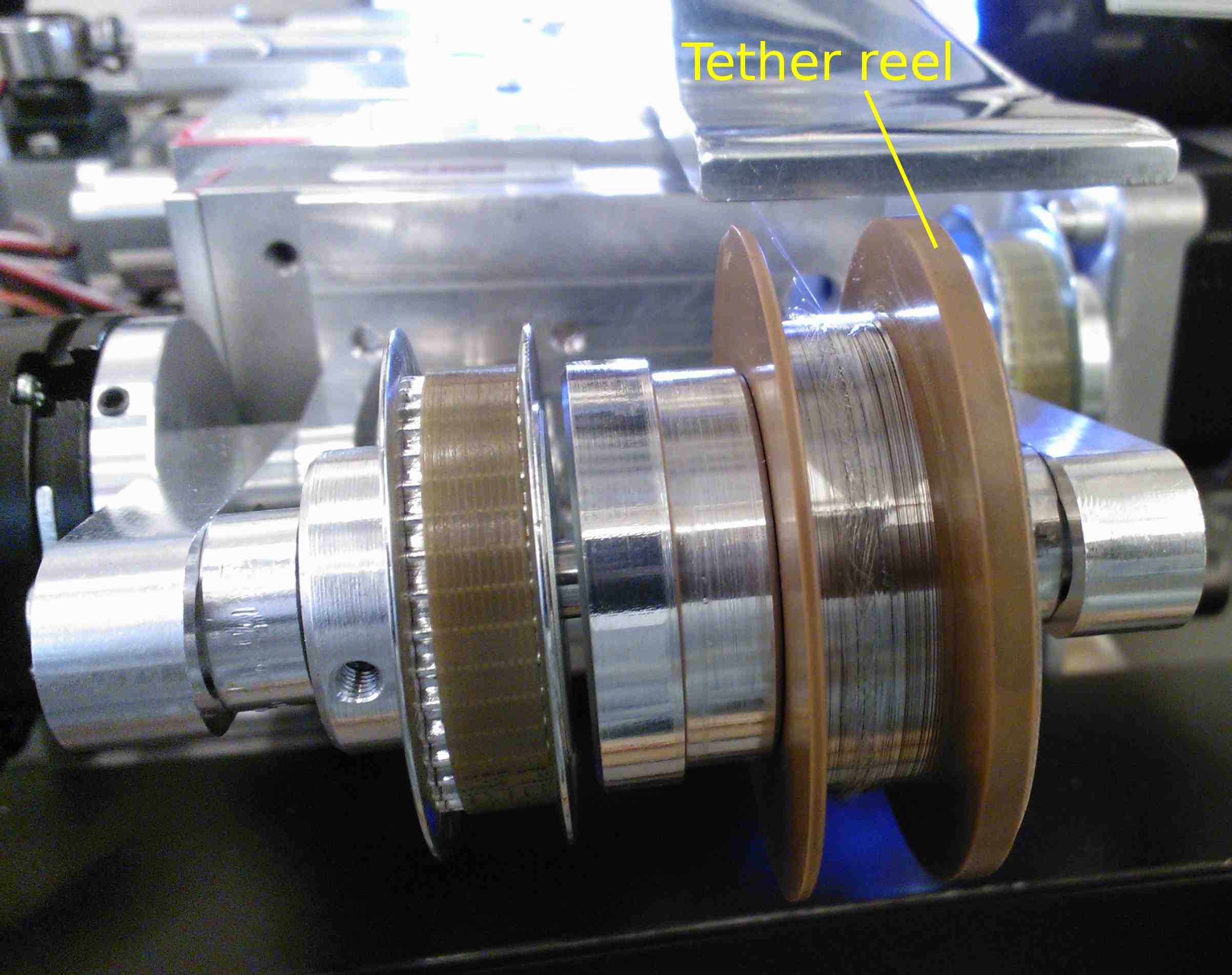}}
\caption{
Tether being manufactured on the reel at UH in January 15th, 2013.
}
\label{Reeling}
\end{figure}

\begin{figure}
\centerline{\includegraphics[width=0.6\columnwidth]{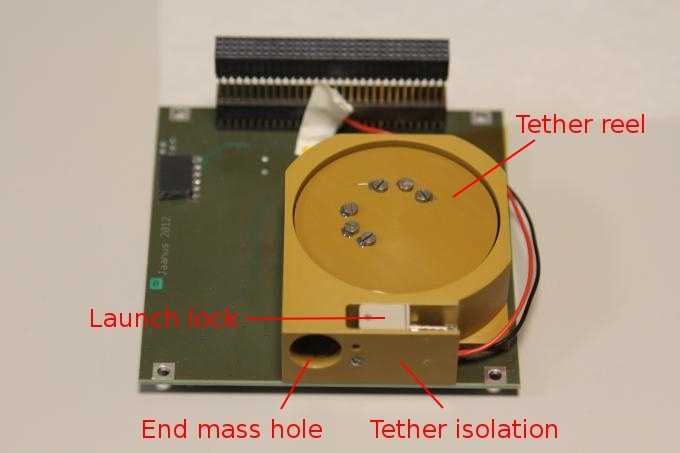}}
\caption{
Motor board partly assembled. The two electric wires on the
right are used for releasing the launch lock. They were not
connected to their terminals on the circuit board at the time of
this photograph.  }
\label{FigPL1}
\end{figure}

In the beginning of the mission the tether is stored on a dedicated
reel. The overall diameter and height of the reel are 50~mm and 14~mm,
respectively. The inner cylindrical face, around which the tether is
stored, has the dimensions of 36~mm and 10~mm for diameter and height,
respectively. Packing the tether on the reel is a delicate process,
which currently can only be performed by the tether factory at UH. The
tether is packed directly onto the target reel while being
manufactured. Figure~\ref{Reeling} shows this work in
progress. Onboard the satellite the reel is attached directly on the
motor which runs it. Surrounding the reel there is a block called
tether isolation. The purpose of this block is to offer mechanical
platform for the tether end mass and its launch lock. Also, the tether
isolation block ensures that the exposed high voltage tether is
electrically isolated from its surroundings. Figure~\ref{FigPL1} shows
the tether isolation mounted around the reel. The material of both the
reel and the tether isolation is Tecasint 4011.\cite{EdelbauerAndPorn}
As a polyimide it is known to have favorable mechanical, thermal and
electrical properties for space use. It also comes with prior flight
heritage. These mechanical parts were designed and produced by the
German Aerospace Center, DLR Bremen.

\subsection{Reel motor and control electronics}
\begin{figure}
  \centering
  \subfloat[][Reel motor attached on the circuit board. Shown circled are the four slip ring contacts.]{\label{FigStripdownA}\includegraphics[width=0.5\columnwidth]{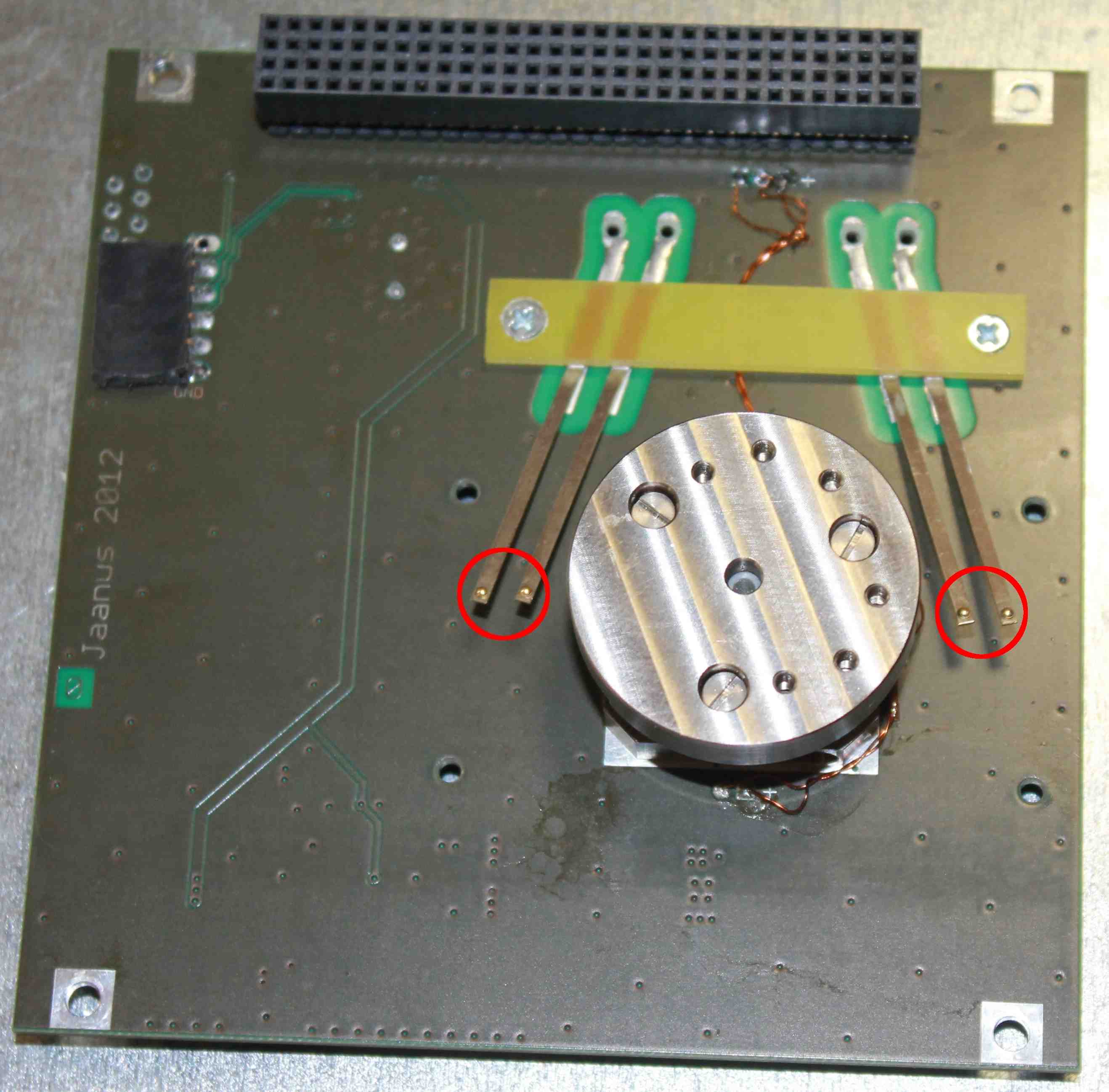}}~
  \subfloat[][Tether reel attached on the motor.]{\label{FigStripdownB}\includegraphics[width=0.5\columnwidth]{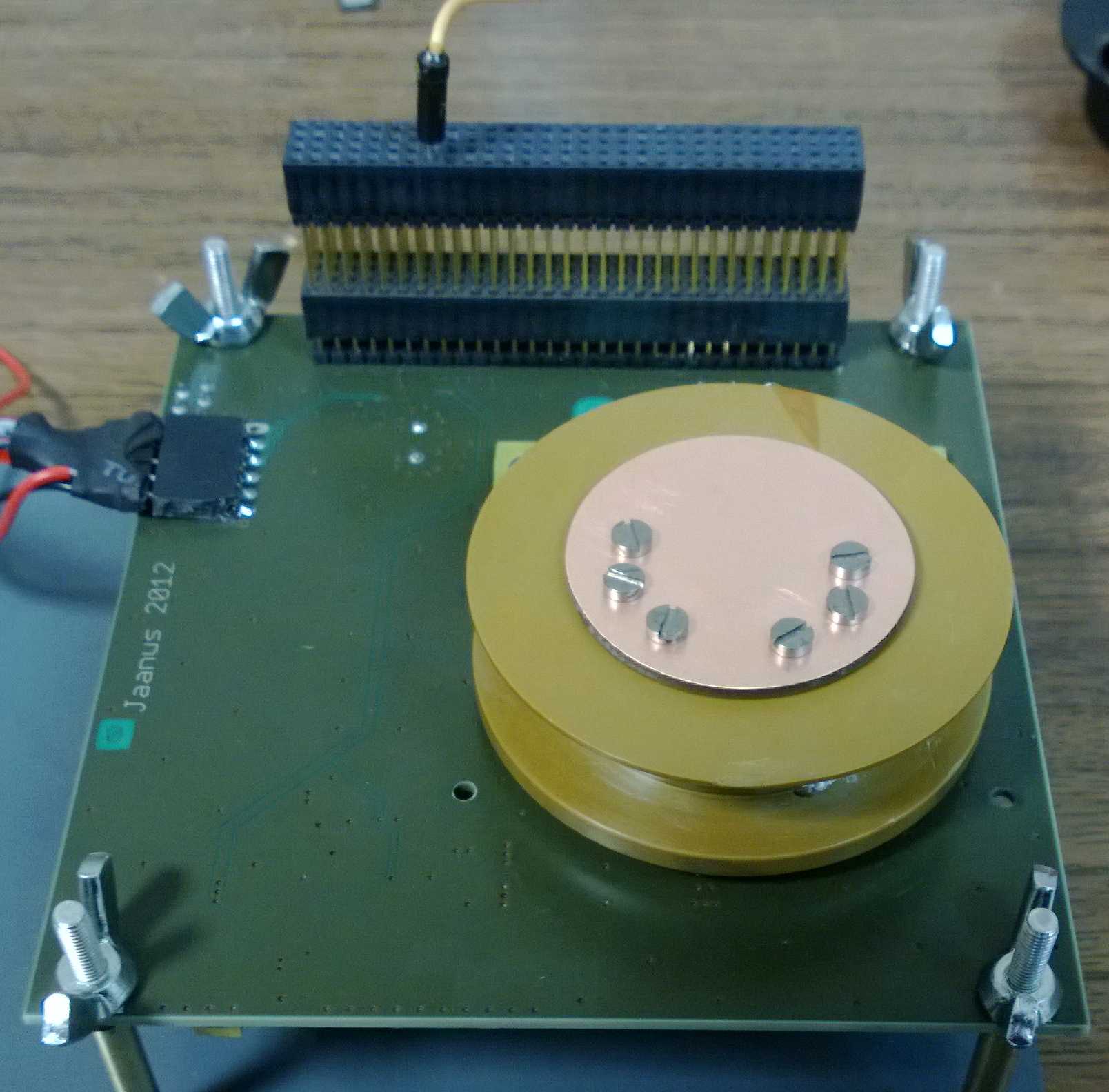}}
  \caption{Different stages of motor board assembly.}
  \label{FigStripdown}
\end{figure}

A closed loop precision piezo rotator ANR101/RES from Attocube
corporation was chosen as the tether reel motor. In
Figure~\ref{FigStripdownA} the motor can be seen attached on the motor
board. The design and dimensions of the motor allowed it to be placed
inside the tether reel, thus saving space on the circuit board, as
shown in Figure~\ref{FigStripdownB}. The motor is controlled with
custom designed control electronics. The design of the control
electronics, as well as the circuit board layout and manufacture, were
carried out by the University of Tartu (UT).

\subsection{Slip ring}
\begin{figure}
  \centering
  \subfloat[][Bare slip ring.]{\label{FigSlipringA}\includegraphics[width=0.5\columnwidth]{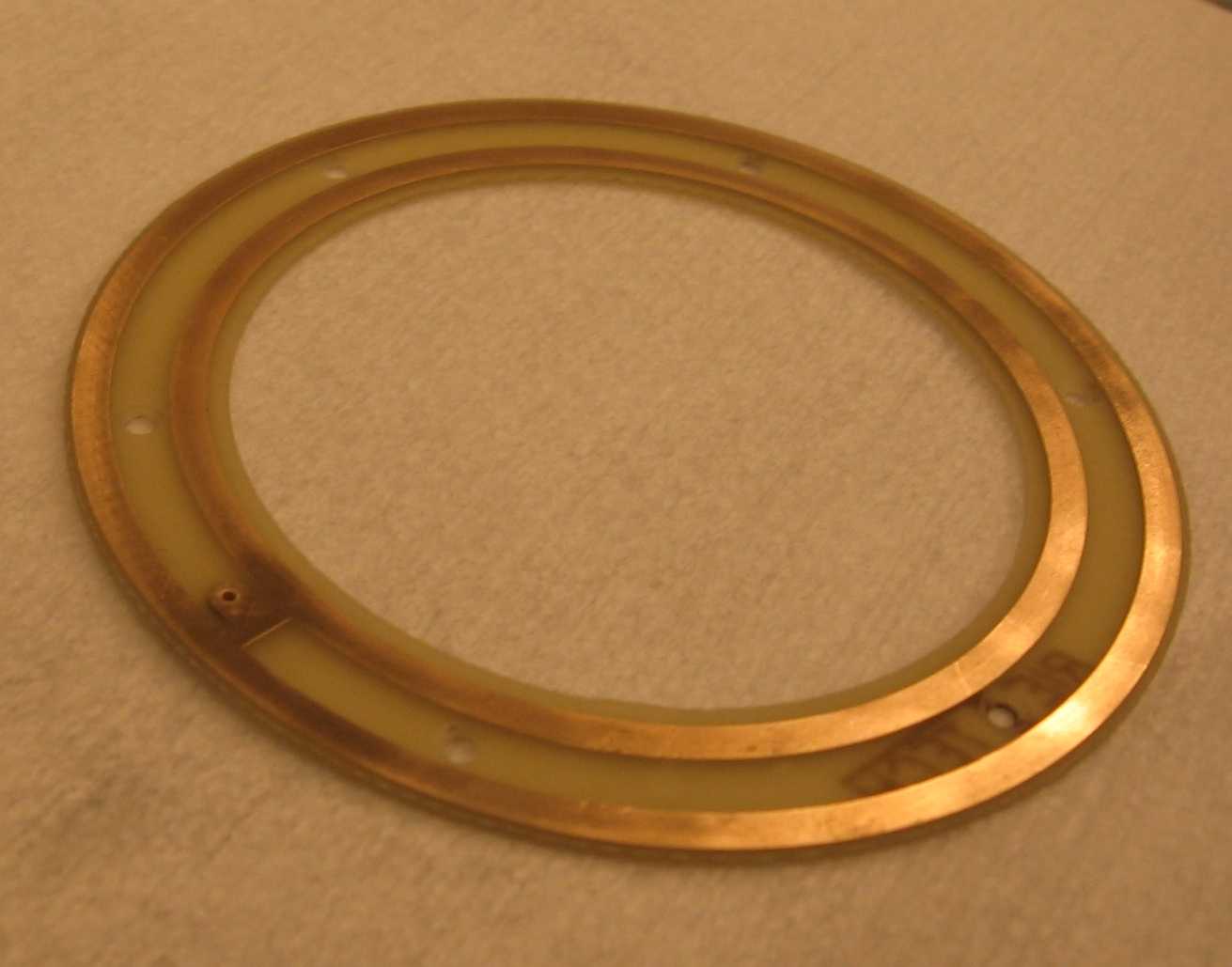}}~
  \subfloat[][Slip ring attached on reel, tether bonded and reeled.]{\label{FigSlipringB}\includegraphics[width=0.5\columnwidth]{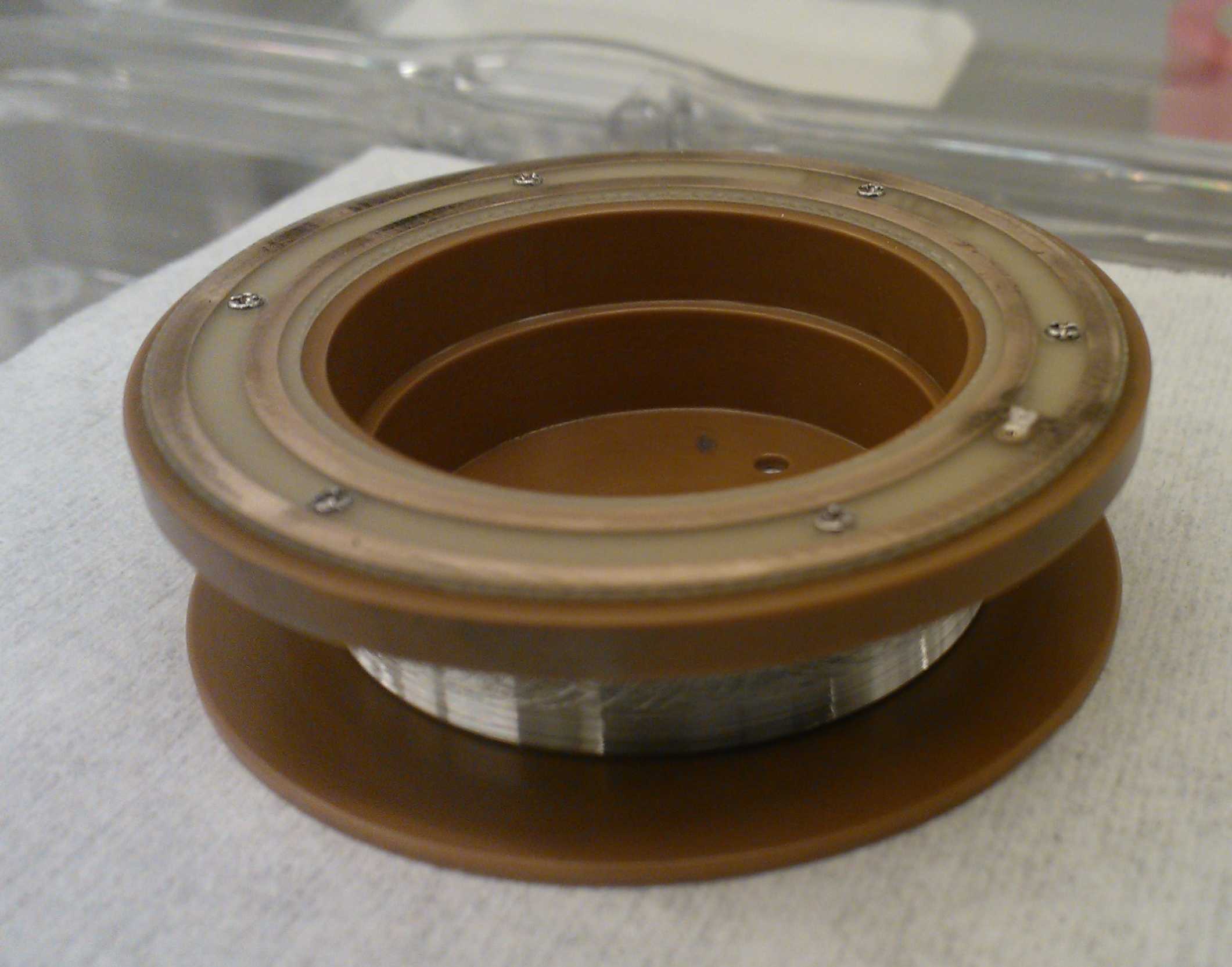}}
  \caption{Slip ring.}
  \label{FigSlipring}
\end{figure}

Because the tether needs to be charged with a high voltage during the
experiment, an electrical contact is needed between the tether and the
HV source. To establish such contact from the circuit board onto a
rotating reel, a slip ring structure is
used. Figure~\ref{FigSlipringA} shows the slip ring. It is constructed
of circuit board material with a gold alloy pattern etched on it. In
Figure~\ref{FigSlipringB} the slip ring is shown attached on the
bottom of the reel with six screws. The four slip ring contacts can be
seen in Figure~\ref{FigStripdownA} connected on the board and
surrounding the motor on both sides. The material of the slip ring
contacts is copper, with the contact points at the tip coated with
gold alloy.

\begin{figure}
  \centering
  \subfloat[][]{\label{FigSlipringBondA}\includegraphics[width=0.5\columnwidth]{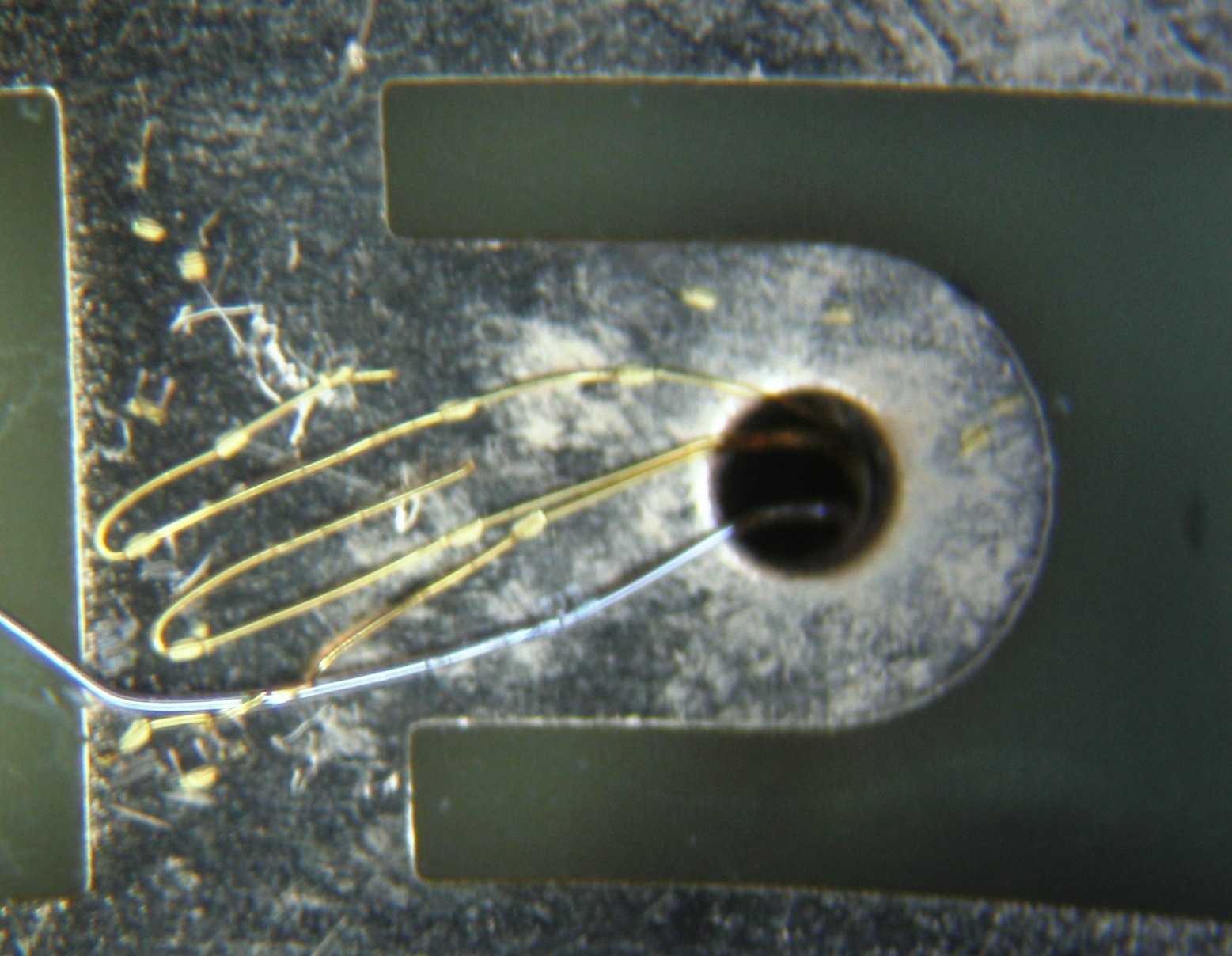}}~ 
  \subfloat[][]{\label{FigSlipringBondC}\includegraphics[width=0.5\columnwidth]{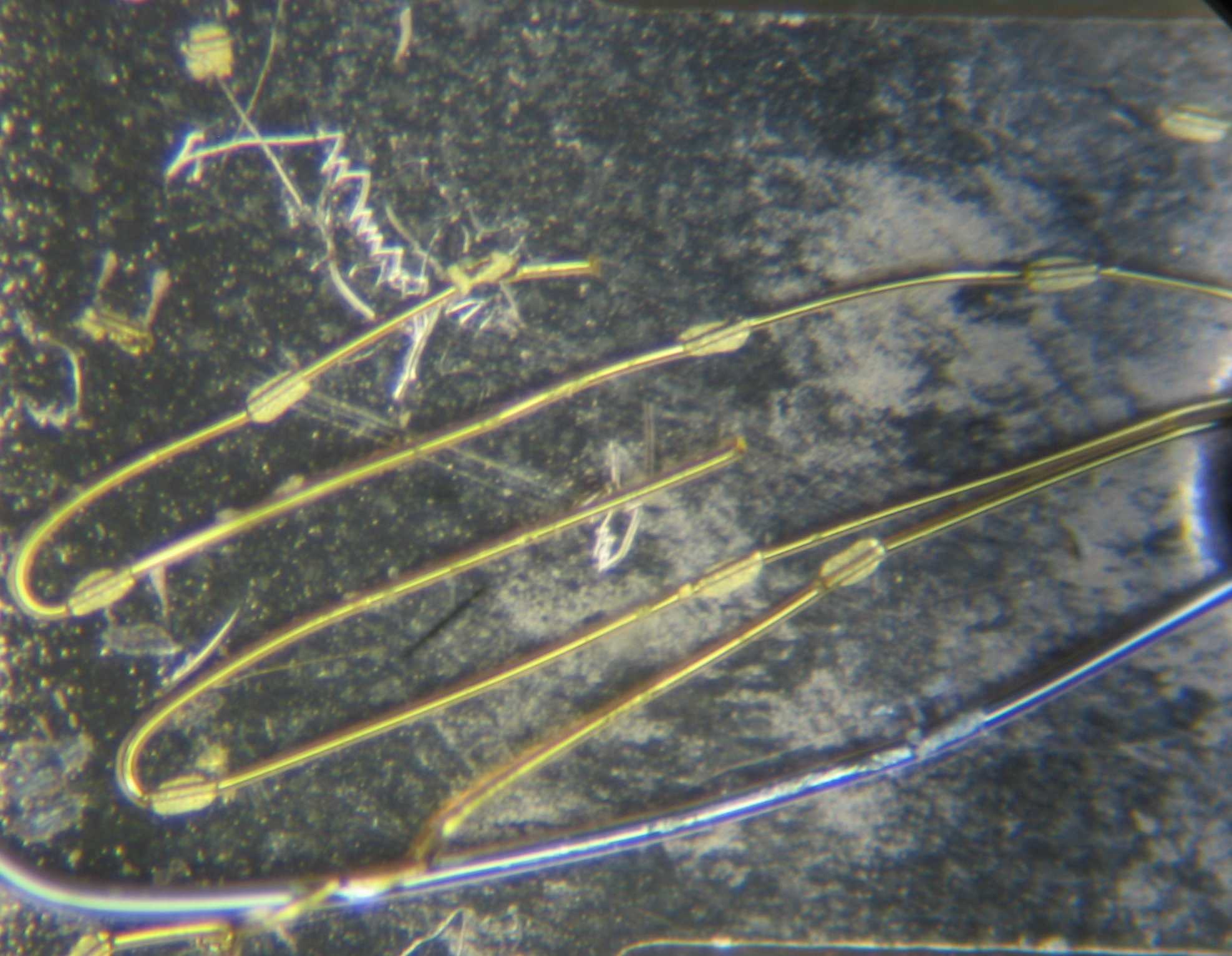}}\\
  \subfloat[][]{\label{FigSlipringBondB}\includegraphics[width=0.5\columnwidth]{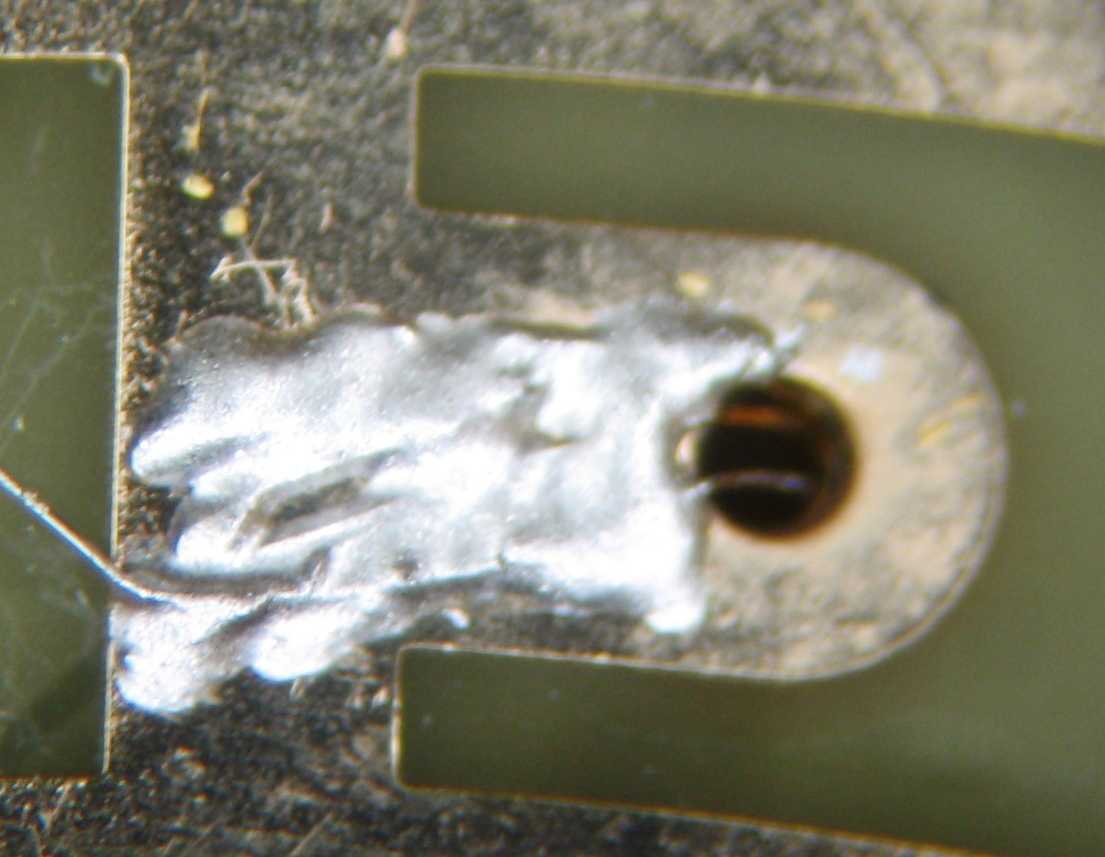}}
  \caption{Slip ring bond. (a) Basewire and three gold wires bonded on
    slip ring. (b) Closeup of the same bonds. (c) All bonds covered
    with conducting glue.}
  \label{FigSlipringBond}
\end{figure}

Figure~\ref{FigSlipringBond} depicts how the tether is contacted to
the slip ring. The root of the tether's basewire emerges through a
hole. Three gold wires had been bonded to the basewire and
subsequently all four wires were bonded to the slip ring. The reason
for adding the gold wires was the difficulty of producing high quality
bonds between the basewire and the slip ring gold alloy. Finally all
bonds were covered with the conducting H20E glue from Epotek to ensure
proper electrical and mechanical contact. The slip ring was designed
and manufactured by DLR, while the tether attachment was handled by
UH.

\subsection{End mass}
\begin{figure}
  \centering
  \subfloat[][]{\label{SubFigEndMassHalves}\includegraphics[width=0.5\columnwidth]{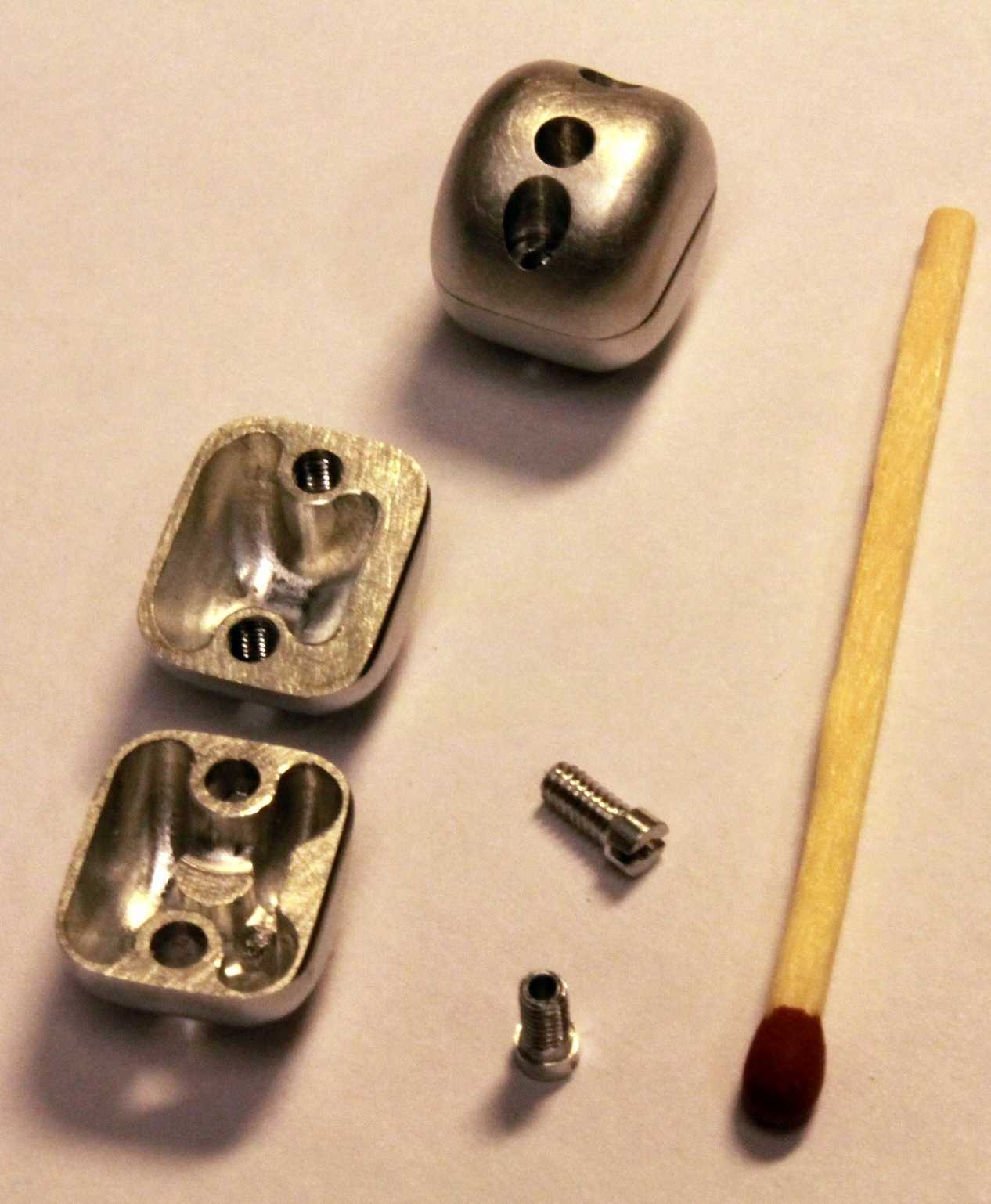}}~
  \subfloat[][]{\label{SubFigEndMass}\includegraphics[width=0.5\columnwidth]{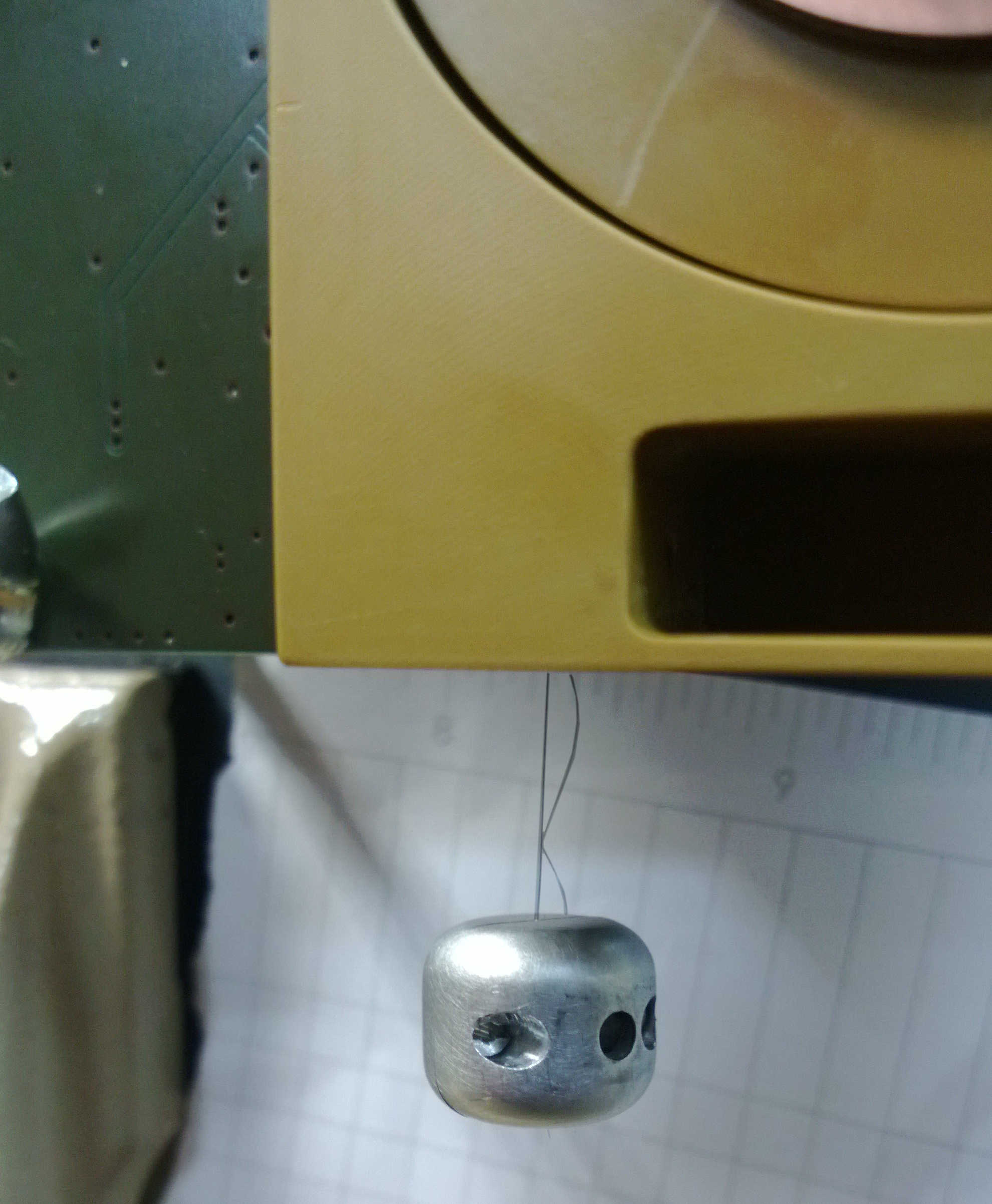}}
  \caption{(a) Assembled and stripped down view of end mass. (b) End
    mass hanging from the tether during PL assembly. Tether enhanced
    digitally for visibility. Notice the middle hole for the launch
    lock pin.}
  \label{FigEndMass}
\end{figure}

As described above, the tether is manufactured of hair thin aluminum
wires. Relying on the mass of the tether alone in an attempt to deploy
it could lead to failure. To increase the centrifugal pull experienced
by the tether, a 1.2 gram end mass, with dimensions of 12~mm and 10~mm
in diameter and length, respectively, was added at the tip. The end
mass is constructed of aluminum and it consists of two halves, which
are attached with screws, see Figure~\ref{SubFigEndMassHalves}. The
tether was squeezed between the two halves. Friction between the
tether and the two halves was found sufficient, so no additional
bonding or glueing was applied. The end mass was designed and produced
at UH. The assembled end mass can be seen hanging from the tether
during PL assembly in Figure~\ref{SubFigEndMass}.

\subsection{Launch lock}
\begin{figure}
  \centering
  \subfloat[][]{\label{FigLaunchlockA}\includegraphics[width=0.5\columnwidth]{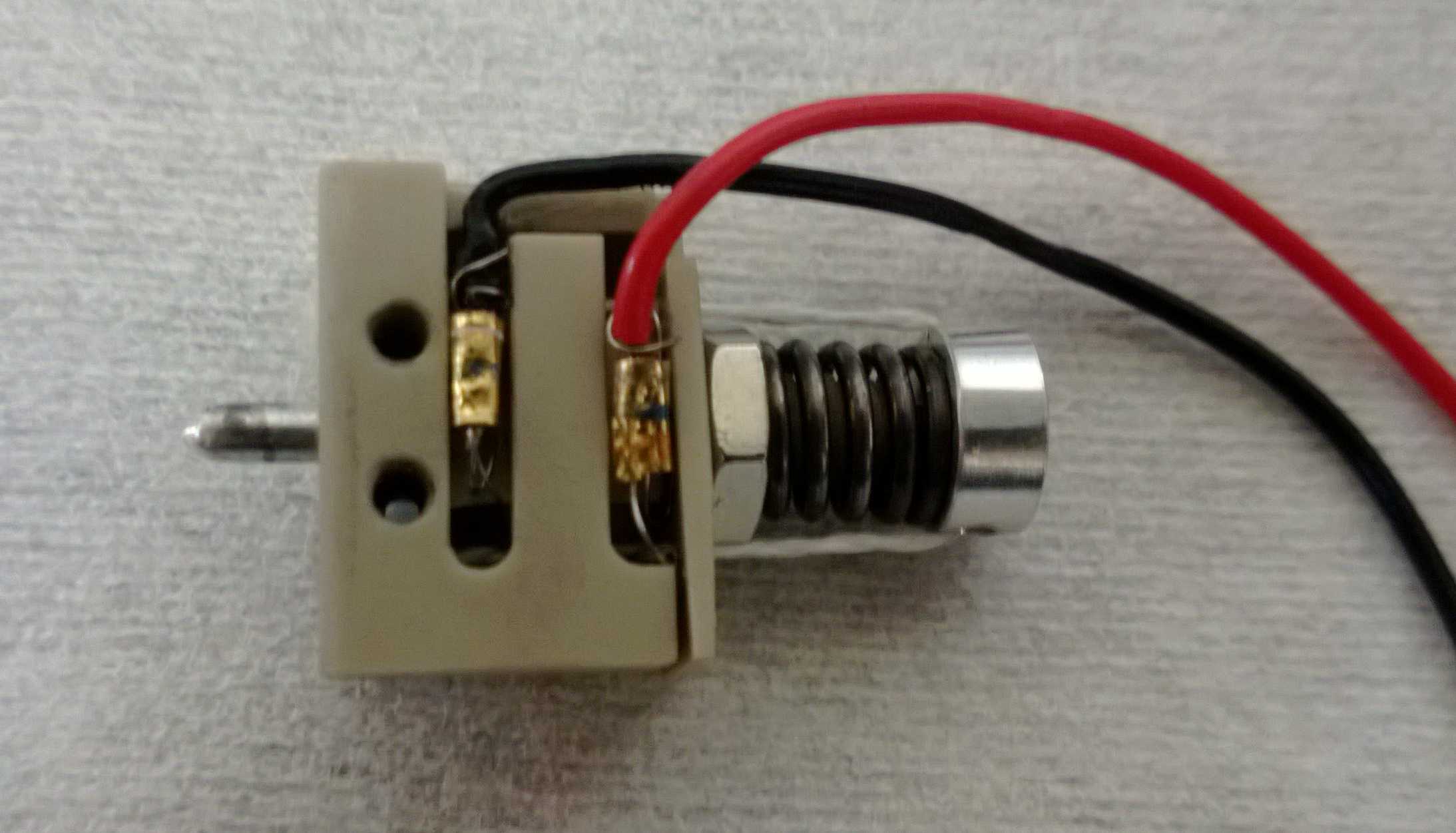}}~
  \subfloat[][]{\label{FigLaunchlockB}\includegraphics[width=0.5\columnwidth]{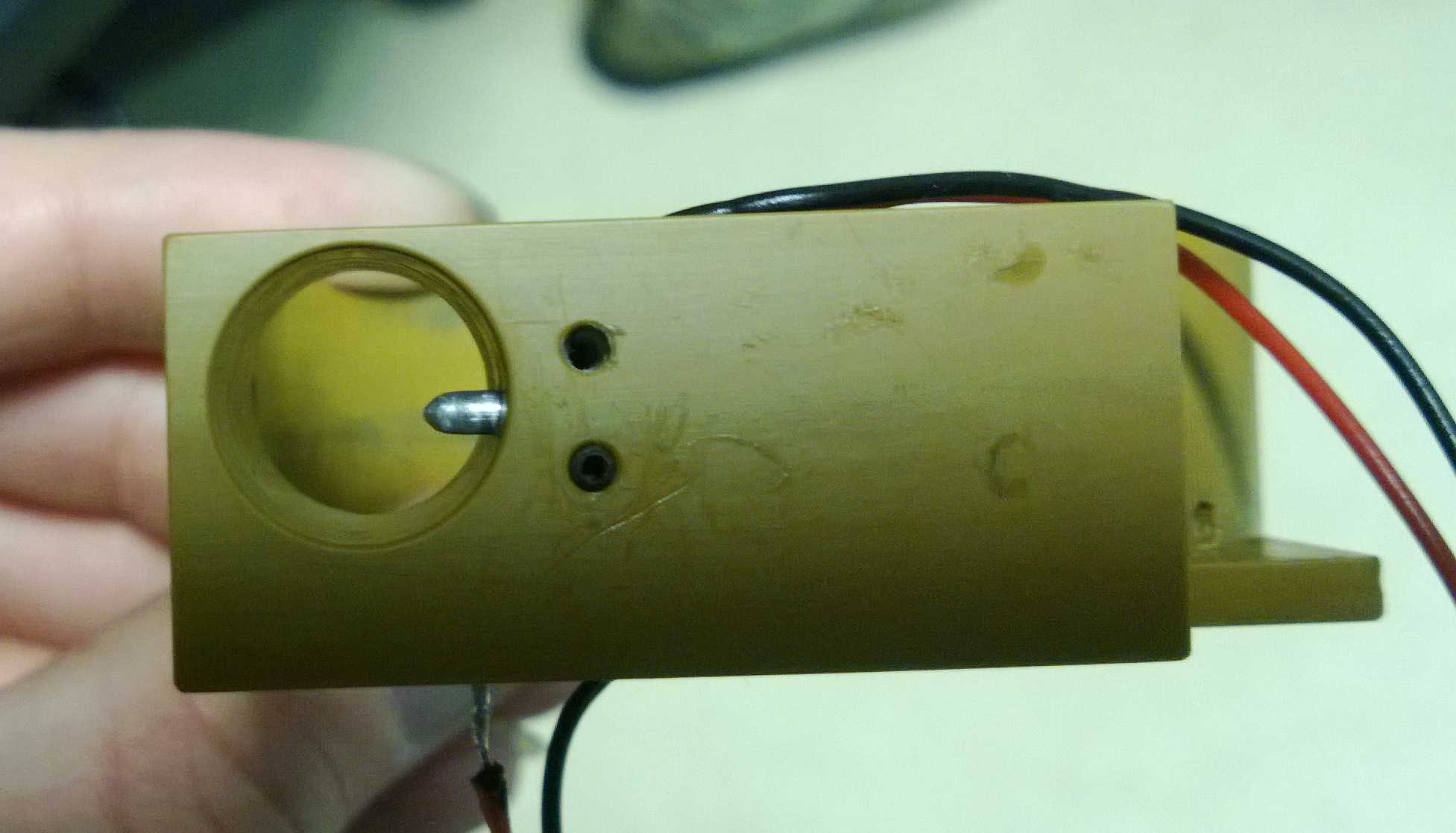}}
  \caption{(a) Launch lock, loaded. (b) Launch lock placed at its position in tether isolation. Notice the pin entering the end mass compartment.}
  \label{FigLaunchlock}
\end{figure}

\begin{figure}
\centerline{\includegraphics[width=0.6\columnwidth]{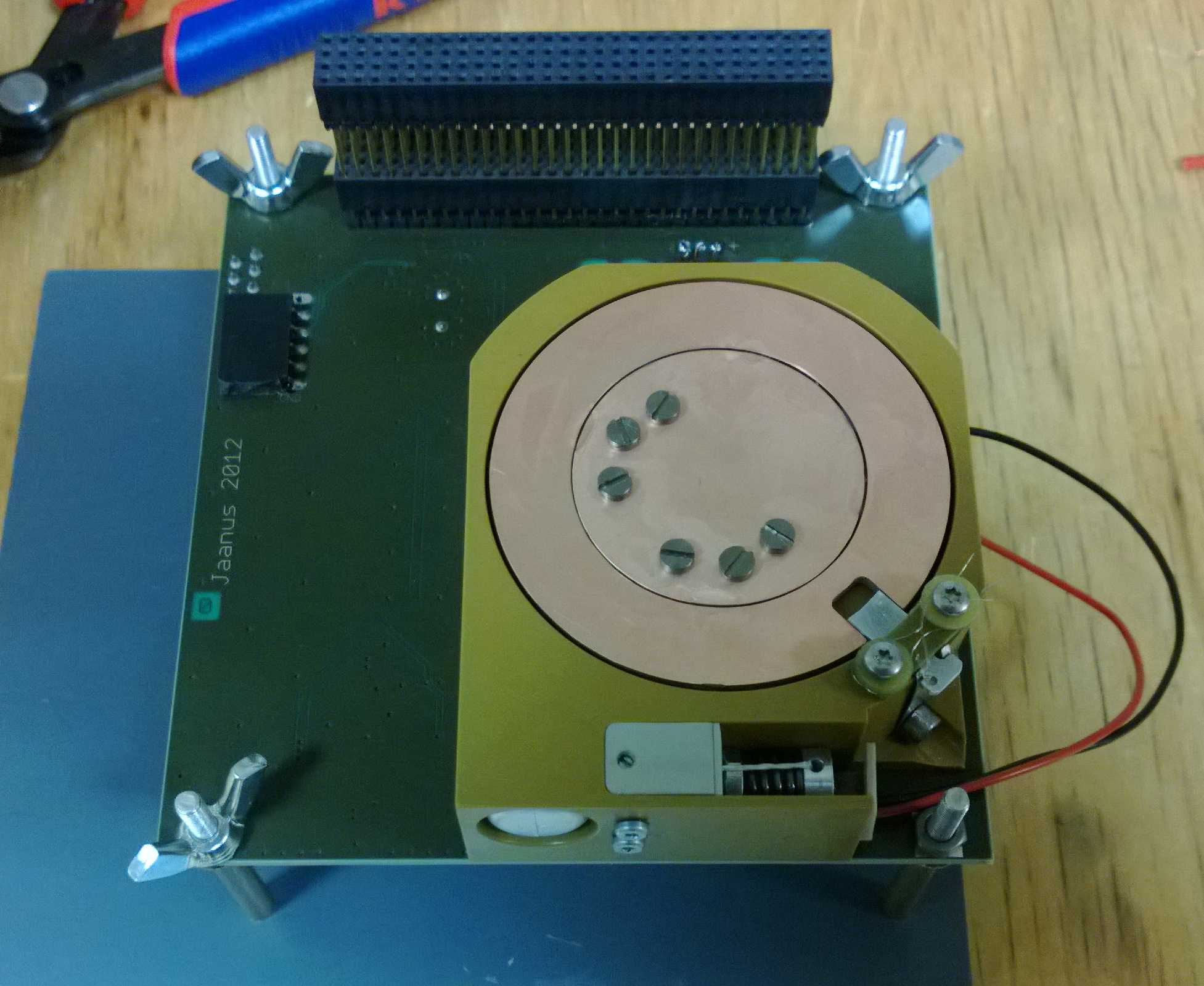}}
\caption{
Motor board fully assembled.
}
\label{FigPL1Fin}
\end{figure}

In the beginning of the mission the end mass must be locked to the
tether isolation. The tensile strength of the tether is only about 50
grams. Unless locked, the end mass would certainly break the tether,
especially during the intense vibrations caused by the launch
rocket. The launch lock can be seen in
Figure~\ref{FigLaunchlockA}. The launch lock design includes a spring
loaded aluminum pin, which enters a dedicated cavity in the end mass
while loaded, see Figure~\ref{FigLaunchlockB}. The spring is kept
loaded with a Dyneema string. Unlocking the reel lock is carried out
by applying a current through a burn wire, which will cut the
string. The launch lock was designed and produced by DLR. The launch
lock can also be seen at its final attached position in
Figure~\ref{FigPL1Fin}.

\subsection{Reel lock}
\label{SecReelLock}
The reel lock is an instrument which needed to be introduced at a
later stage of the satellite manufacture. Originally it was believed
that the friction of the reel motor and its gear system would be
sufficient to prevent the reel from turning during the launch
vibrations. During the first vibration tests this assumption turned
out to be faulty. In particular, it was observed that certain
resonance frequencies during the vibrations would cause the piezo
rotator to turn and thus break the tether. Figure~\ref{FigPL1Fin}
shows the motor board in its final form before integration to the
satellite bus. The reel lock can be seen at the bottom right corner of
the tether isolation block. The locking is based on a metal bar, which
is pushed into a dedicated slot in a locking ring. The locking ring
consists of two coaxial parts, shown attached on top of the tether
reel. The inner part is attaced to the reel with the same screws which
lock the reel to the motor. The purpose of this central part was to
guide the outer part at the center while being glued at its place. The
reason for this complicated arrangement arose from the fact that the
assembly procedure of the motor board required the end mass to be
reeled in at its correct place in the tether isolation block with
sub-millimeter accuracy. This in turn prevented us from predetermining
the place of the locking slot on the face of the reel. The correct
place would be dictated with sub-millimeter precision by the reel-in
process. Therefore, the outer ring of the locking mechanism needed to
be placed at its final position as the very last stage of the
assembly. The unlocking is initiated by use of a burn wire to release
the metal bar, which shall then be retracted by a vertically placed
leaf spring. The design and manufacture of the reel lock was carried
out by UT.

\subsection{High voltage board}
\label{SecHV}
\begin{figure}
\centerline{\includegraphics[width=0.6\columnwidth]{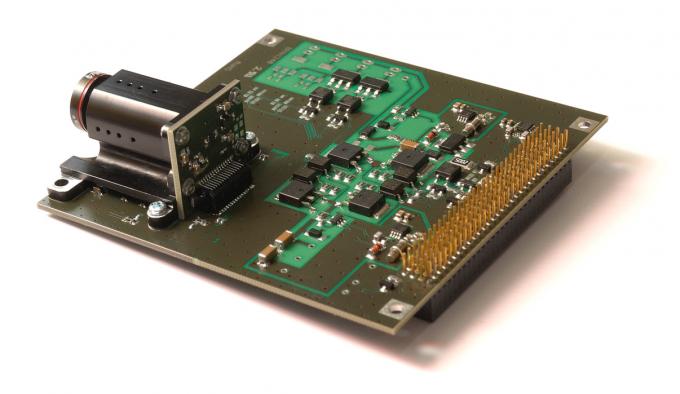}}
\caption{
High voltage board. The large black block is the onboard camera.
}
\label{FigHV}
\end{figure}

The high-voltage source is responsible for generating $\pm$500~V
$\pm$5$\%$ voltage, which shall be used to charge the tether, as well
as powering the electron guns (in positive voltage mode). All of the
components for that are on the high-voltage board in the PL section of
the satellite that is powered by 3.3~V, 5~V and 12~V lines from
EPS. Figure~\ref{FigHV} shows the finished board.

\begin{figure}
\centerline{\includegraphics[width=\columnwidth]{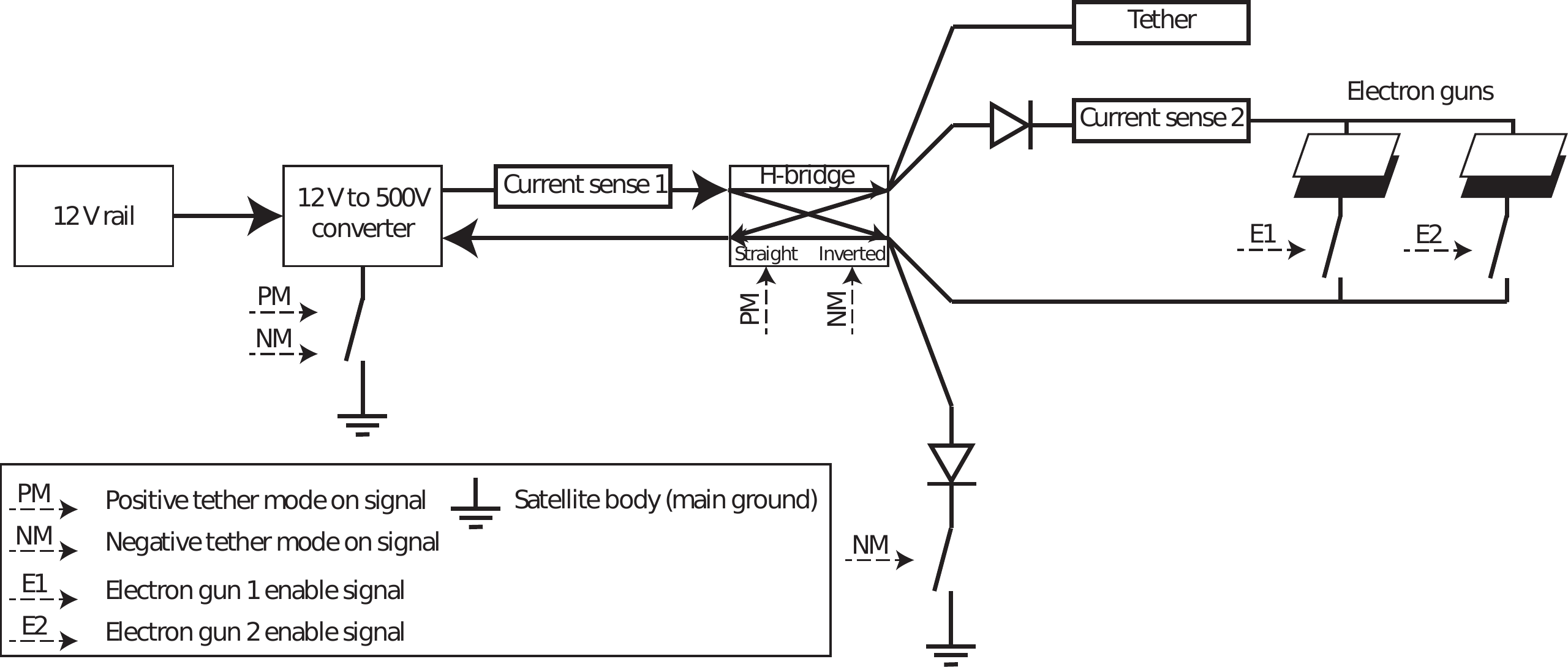}}
\caption{
Schematic presentation of the HV supply.
}
\label{FigHVSchema}
\end{figure}

The basic principle (see Figure~\ref{FigHVSchema}) is based on using
an isolated 12~V to 500~V DC-DC converter (12SA500 from Pico
Electronics), output of which is connected through an H-bridge between
the body of the satellite, the electron guns and the tether. Currents
going into the tether and the electron guns can be monitored. The
output voltage can be changed by modifying the 12~V rail voltage by
$\pm$5$\%$, since the output voltage of the converter depends on the
input voltage proportionally.

In positive tether mode, the H-bridge connects the positive output of
the converter to the tether and to the anodes of the electron guns. In
this mode the electron guns can be turned on by connecting or
disconnecting the cathodes. In this case electrons are emitted through
the electron guns and are absorbed by the tether.

In negative tether mode, the H-bridge connects the negative output of
the converter to the tether and the positive output is connected to
the satellite body. In this case the positive ions are being gathered
by the tether and electrons are being absorbed by the satellite body
that is connected to the main ground connections of the satellite.

A challenge of this design is that nothing on the high-side of the
500~V converter is referenced to the satellite ground and very high
voltages are involved. This especially complicates current sensing. In
our approach we referenced both of the current senses to the negative
output of the 500~V converter. This can be done since the current
sense measuring the anode current is measuring only if it is connected
through the H-bridge directly to the positive output of the 500 V
input. Still, to be able to measure that current with electronics
onboard the satellite, this needs to be re-referenced to the satellite
ground. This was accomplished by using voltage-to-frequency converters
on the 500 V side and transmitting the frequency signal to the
low-voltage side using capacitive coupling. The voltage-to-frequency
converters require a 5 V isolated voltage source on the 500~V side,
since doing down-conversion from 500~V would introduce excessive
losses. On low-voltage side this frequency is converted back into
voltage and it is referenced to the satellite ground. Further an
onboard analog-to-digital converter (ADC) is used so that the CDHS can
read the voltage on the board over Serial Peripheral Interface (SPI)
bus.

Switching is also an issue on the 500 V side, since the command
signals are also not referenced to a constant ground. This issue was
solved by using optotriacs in the H-bridge to drive the bipolar
junction transistors that do the switching. The electron gun cathodes
are switched using n-channel MOSFETs that are driven by photovoltaic
MOSFET drivers (APV2121 from Panasonic). This photovoltaic driver is
especially useful since it can drive n-channel MOSFETs on the low
side. The photovoltaic driver is also used to connect the positive
terminal of the high voltage source to the body of the satellite in
negative tether mode.

As a safety feature the measured current is compared to a reference
value and if it is exceeded the H-bridge is turned off.

\subsection{Electron guns}
The E-sail effect can be observed with either positive or negative
tether voltage. In either case there shall be a flow of charge
carriers flowing from the surrounding plasma to the tether, attempting
to neutralize the tether potential. This flow needs to be compensated
with another flow, with opposite direction of electrical current
between the spacecraft body and the surrounding space. In the case of
ESTCube-1 mission in low Earth orbit the case of the negative voltage
is simple from the apparatus point of view. It has been estimated that
in the plasma conditions of the orbital altitude the rate at which the
satellite body is able to attract electrons from the local plasma
environment would be sufficient to compensate for the rate at which
the tether attracts positive ions. The case is more complicated when
applying the positive voltage. Because of the lower thermal speed of
ions the ion flow to the satellite body is insufficient to compensate
for the electron flow to the tether. Due to this there exists a need
for a way to remove excess electrons from the system in order to
maintain the positive tether potential. For this task we use two
onboard cold cathode electron guns, placed on the top and the bottom
sides, i.e. $\pm$Z sides, of the satellite. The electron guns have
been developed for our mission at the University of Jyv\"askyl\"a
(UJ).

\begin{figure}
\centerline{\includegraphics[width=\columnwidth]{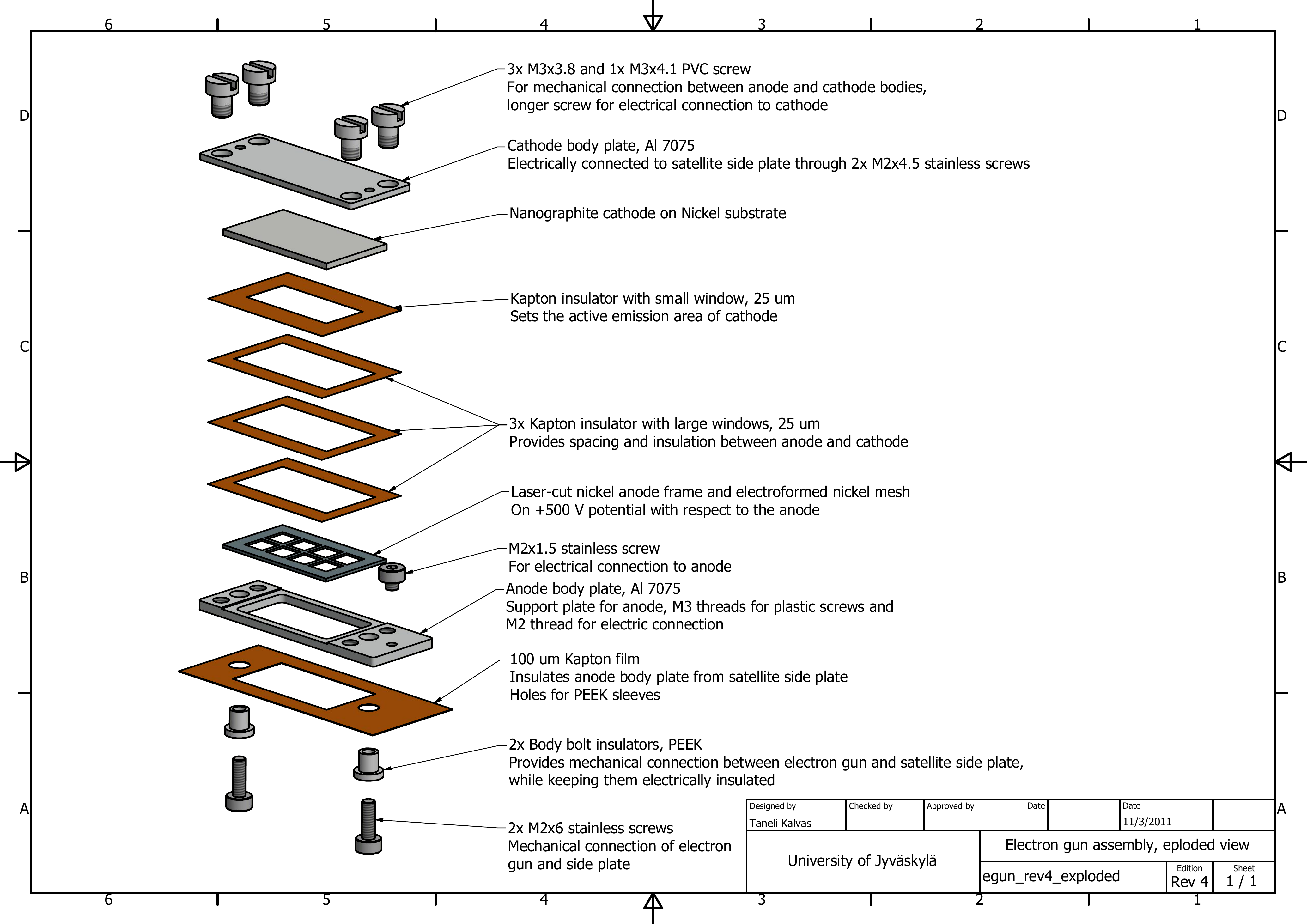}}
\caption{
Exploded view of electron gun.
}
\label{FigEgunExp}
\end{figure}

\begin{figure}
  \centering
  \subfloat[][Top view.]{\label{FigEgunTop}\includegraphics[width=0.6\columnwidth]{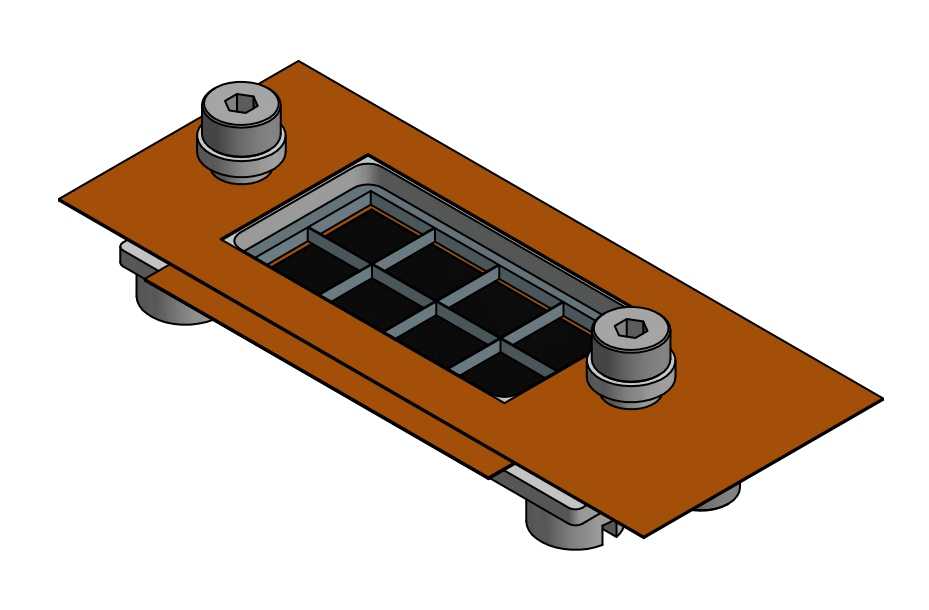}}\\
  \subfloat[][Bottom view.]{\label{FigEgunBottom}\includegraphics[width=0.6\columnwidth]{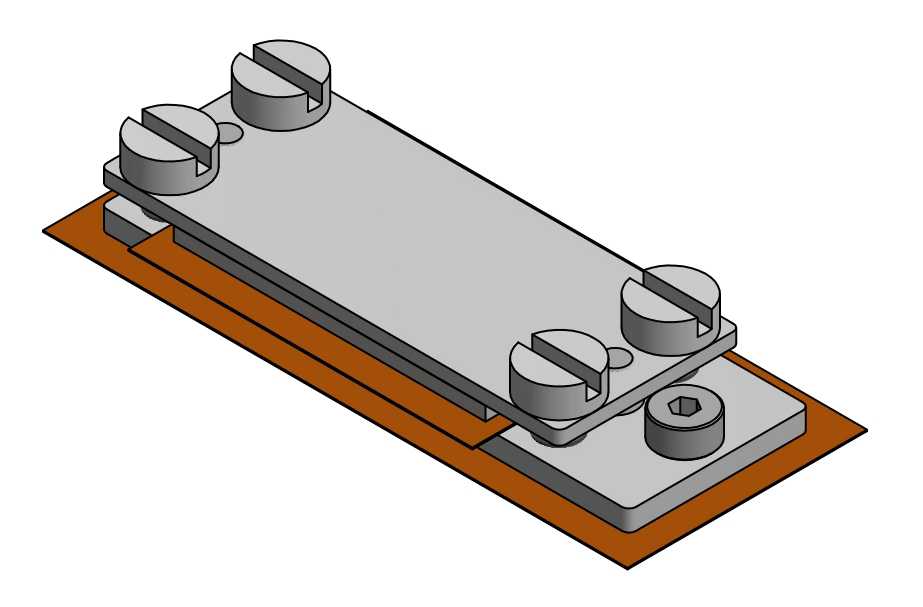}}
  \caption{Two views of the assembled electron gun. "Top" is the direction of the emitted electron beam.}
  \label{FigEgunTopBottom}
\end{figure}

\begin{figure}
\centerline{\includegraphics[width=0.6\columnwidth]{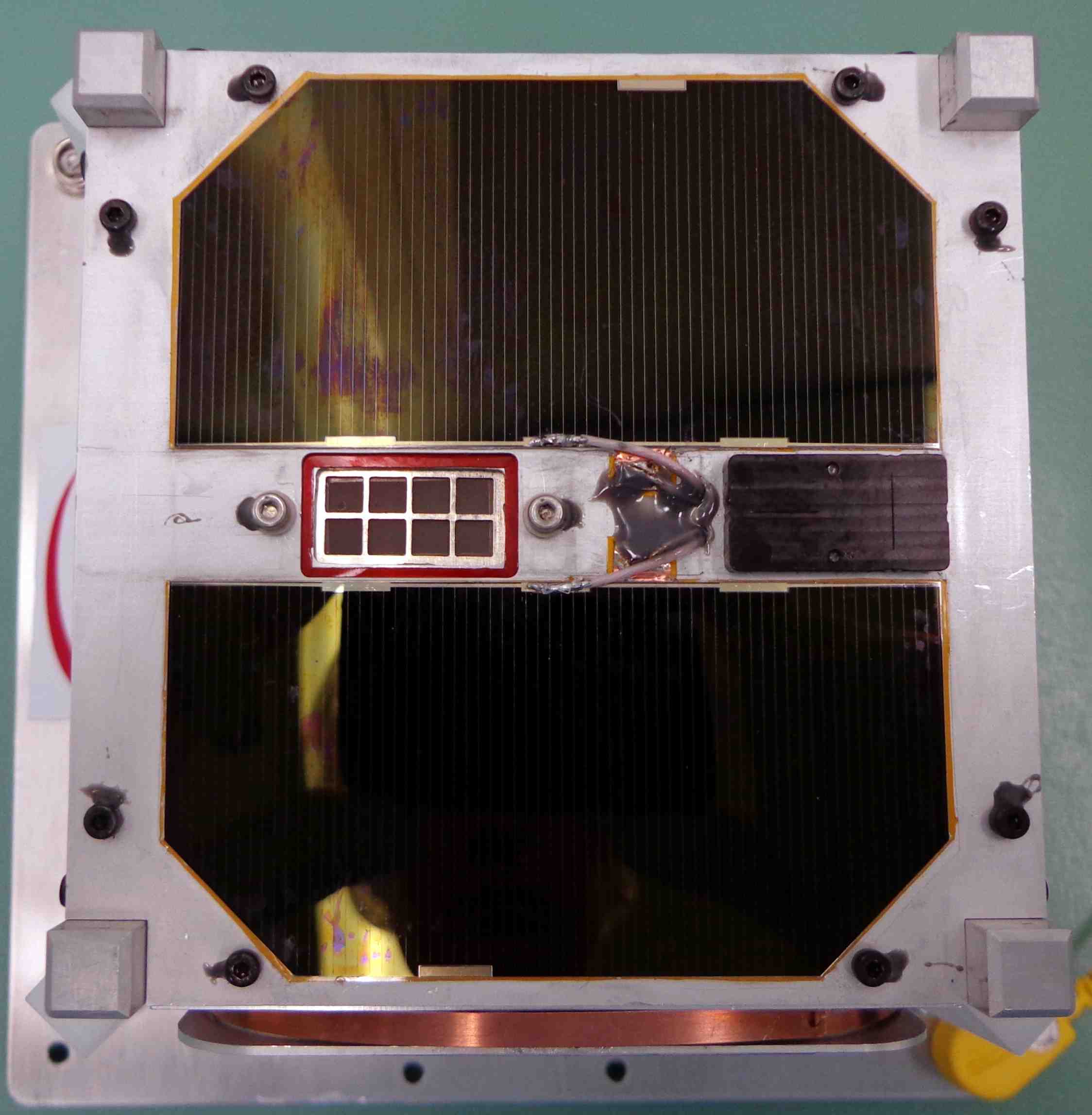}}
\caption{
Electron gun attached on satellite side panel.
}
\label{FigEgunSidepanel}
\end{figure}

Figure~\ref{FigEgunExp} shows the stucture of the electron gun. The
cathode \cite{Obraztsov2009} is made of graphite coated nickel
substrate and the anode is made of a fine electroformed nickel mesh
bonded to a laser-cut nickel frame. The gap between the anode and the
cathode is of the order of 100~$\mu$m. Typical values for the tether
current during the E-sail experiment are 1~mA in the positive mode and
10~$\mu$A in the negative mode. Figure~\ref{FigEgunTopBottom} shows
drawings depicting two views of an assembled electron
gun. Figure~\ref{FigEgunSidepanel} shows an electron gun attached to
the satellite's side panel.

\section{Operations}
The principle of determining the E-sail force, acting between the
charged tether and the ionospheric plasma, is based on observing the
change in the satellite's spin rate as the voltage of the tether is
cycled on and off with correct synchronization with the satellite's
attitude and orbital position. The physics of this procedure are
described with more detail in Ref. \cite{Silver2014}. The exact
sequence of events for the tether experiment is difficult to plan
beforehand, as it is influenced by several factors related to orbital
conditions and general behavior of the spacecraft, which must be
determined at the time of the actual experiment. The baseline sequence
of events is listed in the following.

\begin{enumerate}
\item Spin-up of satellite.
\item Reel lock deployment.
\item Launch lock deployment.
\item Tether deployment, imaging of tether and end mass.
\item Apply negative voltage to tether.
\item Observe change in satellite's spin rate.
\item Apply positive voltage to tether.
\item Observe change in satellite's spin rate.
\end{enumerate}

The tether experiment phase of the mission is begun by ramping up the
satellite's spin rate to approximately 1 rev/s. The spin-up is
operated by the satellite's Attitude Determination and Control System
(ADCS). A more detailed description of the requirements and methods of
the spin-up are given in Refs
\cite{SlavinskisProcEstAcadSci2014,SlavinskisADCS2014,Silver2014}. Once
the proper spin rate and axis have been obtained the reel lock and
launch lock are deployed. During and immediately after the launch lock
deployment the onboard camera is used to take several pictures with
sub-second intervals. This is used as a precaution in case the launch
vibrations have broken the tether, in which case the images should
show the slowly escaping end mass. During the tether deploymet the
motor is used to perform a controlled reel-out with predetermined
speed. Due to its structure the tether cannot be retracted back onto
the reel, the tether deployment is thus an irreversible event. The
tether deployment will most probably be performed in parts. This gives
the ground control team the opportunity to monitor the satellite's
behavior and correct all possible deviations from normal operating
state. Throughout the tether deployment the satellite's spin rate is
monitored and the data are transmitted down to ground station. Also
the tether and the end mass are constantly imaged by the onboard
camera in order to obtain an independent verification of a successful
deploymet. The E-sail force is observed, as described in
\cite{Silver2014}, with both positive and negative voltages. The
negative mode will be tested first, because it does not require the
use of electron guns and is therefore considered less risky. The
essential data for the E-sail force determination are the change in
spin rate and the measured tether current, the latter giving valuable
information about the local plasma environment.

\section{Conclusions and Discussion}

This paper has described the contents of the ESTCube-1 nanosatellite
payload. The payload is used for carrying out the first space
experiment of the E-sail concept.

The equipment was built and delivered to the ESTCube-1 nanosatellite
team in time. The satellite was integrated to the European Space
Agency's (ESA) Vega launch vehicle and the satellite was launched and
delivered into orbit on May 7th, 2013 (UTC) from Europe's Spaceport in
Kourou, French Guiana.

The first vibration tests revealed a design flaw in the motor
board. The piezo rotator was sensitive to certain resonance
frequencies of the vibrations. This caused the tether reel to turn and
thus break the tether. This flaw was subsequently fixed by introducing
the reel lock. The reel lock itself went through some vibration
testing, but the whole tether reel system did not. However, at this
point we already know almost with full certainty that the reel lock
has accomplished its locking duties. A reel lock failure, or a tether
breakage of any sort, would almost certainly lead to the loose tether
filling the insides of the satellite, creating short cuts at several
places. During its time in orbit the satellite has mostly functioned
as expected. This does not support the notion of any launch failures
having taken place. Also, the electron guns were not tested for
vibrations. This was a planned strategy. Because of the very early
stage of development of the miniaturized cold cathode electron gun
technology, only flight models were available within the schedule of
our mission. However, unlike the tether and its related equipment, the
electron guns are not mission critical instruments. Even with no
functional electron guns the negative voltage mode can still be used
for the E-sail experiment.

In addition to the matters discussed above, no particular adversities
have emerged. The systems worked flawlessly in functional ground
tests. At the time of the writing (June 2013) the project team in
Tartu is validating the operations of the satellite in orbital
conditions. Also the final flight software is being written. Schedule
for the E-sail tether experiment has not yet been fixed. Once made,
the experiment is also expected to provide some useful data for our
next mission. In 2014 or 2015 it is time to launch the first Finnish
satellite, Aalto-1. It is a 3U CubeSat designed and built by the
students of Aalto University, Espoo, Finland. A similar tether payload
will be included in Aalto-1 as one of its secondary payloads. The
length of the tether in Aalto-1 will be 100 meters.

{\it Acknowledgements\ }Following persons are acknowledged for their
contribution to this work. Jouni Polkko and Sini Merikallio from
FMI. Risto Kurppa, Tuomo Ylitalo and G\"oran Maconi from UH. Olaf
Kr\"omer and Roland Rosta from DLR. Pekka Salminen from SkyTron
co. Alexander Obraztsov from the University of Eastern Finland
Joensuu.

\section{List of Acronyms}
\begin{itemize}
\item ADC~---~Analog to Digital
\item ADCS~---~Attitude Determination and Control System
\item CDHS~---~Command and Data Handling System
\item EPS~---~Electrical Power System
\item ESA~---~European Space Agency
\item FMI~---~Finnish Meteorological Insitute
\item DLR~---~German Aerospace Center
\item HV~---~High Voltage
\item MCU~---~Microcontroller Unit
\item MOSFET~---~Metal Oxide Semiconductor, Field Effect Transistor
\item PL~---~Payload
\item SPI~---~Serial Peripheral Interface
\item UH~---~University of Helsinki
\item UJ~---~University of Jyv\"askyl\"a
\item UT~---~University of Tartu
\end{itemize}

\bibliographystyle{vancouver-proceedings}
\bibliography{esail}

\clearpage

\clearpage

\clearpage

\clearpage

\clearpage

\clearpage

\clearpage

\clearpage

\clearpage

\clearpage

\clearpage

\clearpage

\clearpage

\clearpage

\clearpage

\clearpage

\end{document}